# Observation of Accelerating Wave Packets in Curved Space

Anatoly Patsyk,[1] Miguel A. Bandres,[1] Rivka Bekenstein,[2,3] and Mordechai Segev[1]

[1]*Physics Department and Solid State Institute, Technion, 3200003 Haifa, Israel*
[2]*Physics Department, Harvard University, Cambridge, Massachusetts 02138, USA*
[3]*ITAMP, Harvard-Smithsonian Center for Astrophysics, Cambridge, Massachusetts 02138, USA*

We present the first experimental observation of accelerating beams in curved space. More specifically, we demonstrate, experimentally and theoretically, shape-preserving accelerating beams propagating on spherical surfaces: closed-form solutions of the wave equation manifesting nongeodesic self-similar evolution. Unlike accelerating beams in flat space, these wave packets change their acceleration trajectory due to the interplay between interference effects and the space curvature, and they focus and defocus periodically due to the spatial curvature of the medium in which they propagate.

In 1979, Berry introduced a unique solution to the potential-free Schrödinger equation: a propagation-invariant wave function, shaped as an Airy function, that accelerates in time in a shape-preserving fashion [1]. This idea remained dormant for almost 30 years, until Christodoulides and his team demonstrated, theoretically and experimentally [2,3], the existence of an accelerating electromagnetic (EM) wave packet: the Airy beam—a paraxial beam that propagates along a parabolic trajectory while preserving its shape. Since the Airy function is not square integrable, it carries infinite power and ideally such beams should be launched from an infinite aperture. Physically, as shown in Ref. [3], truncated Airy beams still exhibit shape-preserving acceleration but for a finite distance only. These results were subsequently generalized to two dimensions, where additional solutions, accelerating parabolic beams, are also possible [4,5]. However, since the Airy beams are exact solutions of the paraxial wave equation, their shape-invariant acceleration is limited to small angles, and their features must be much larger than the optical wavelength. This limitation pushed researchers to search for beams that have nonparaxial features and can accelerate to large angles. Indeed, accelerating beams that are exact solutions of Maxwell's equations were discovered in 2012 [6] and observed soon thereafter [7–10]. These beams follow circular trajectories and can exhibit shape-invariant acceleration up to angles as large as 180° (asymptotically). Shortly thereafter, nonparaxial accelerating beams were generalized to new families in two and three dimensions [11–15]. All these discoveries opened a new route in optics and were followed by additional ideas of accelerating beams in arbitrary convex trajectories [16,17], accelerating beams in photonic crystals [18–20], and in nonlinear media [21–24].

The main feature of all accelerating beams carrying finite power is that, although their center of mass follows a straight trajectory—as expected from conservation of momentum—their main lobe follows a curved path in a shape-preserving manner, up to large distances. This has major consequences, since all light-matter interactions depend on the local intensity of the beam, not on the center of mass (which is an average quantity). For example, the optical forces of radiation pressure and the gradient force depend on the local intensity; likewise, stimulated emission, ionization, and nonlinear optics effects such as frequency conversion and self-focusing all depend on the local intensity. Thus, even though the "center of mass" of the finite accelerating beams moves on a straight line, still, their accelerating main lobe gives rise to effects occurring on curved trajectories. Naturally, this enabled a variety of applications ranging from curved plasma channels using femtosecond Airy beams [25], manipulation of microparticles in nonconventional ways [26–28], and laser micromachining along a curve [29], to single-molecule imaging using the curved point-spread function [30], light-sheet microscopy using Airy beams [31], among many others. In recent years, it became clear that the concept of accelerating wave packets does not belong only to optics; rather, it is a general phenomenon that appears in any wave system. For example, accelerating wave packets were introduced in sound waves [32,33], electron beams [34], and even for relativistic fermions [35]. More recently, the concept of accelerating beams was proposed into curved-space optics [36]. This paper presents the first experimental observation of shape-preserving accelerating wave packets in curved space.

Let us briefly explain the ideas underlying optics in curved space. Perhaps the best-known example of electromagnetism in curved space comes from general relativity, where light is propagating in the presence of the curved space in the proximity of a massive object, which bends the trajectory of the light. However, gravitational effects are extremely weak in a laboratory setting. On the other hand, research on optics in curved-space enables curved-space experiments in the lab, by utilizing the equivalence between the curved-space landscape for EM waves induced via GR effects and the curved space created by space-dependent electric and magnetic perambilities [37]. Most certainly, it is intriguing to have the ability to observe curved-space phenomena in the laboratory in a controllable system, where the physical parameters can be modified at will. Exactly for this cause various experimental systems suggest analogous phenomena that can be demonstrated in laboratory experiments, ranging from cold atoms [38–41] to optical systems [42–46] and metamaterials [47]. One example is using a moving dielectric medium that acts as an effective gravitational field on the light [44]. Another route for such studies is to create curved space by engineering the geometry of the space itself. This idea was first demonstrated by Batz and Peschel [48], who explored the dynamics of light propagating within a thin-film waveguide covering the curved surface area of a three-dimensional body [49,50], effectively creating 2D curved space for the light. However, thus far, in all of these experiments on curved-space optics, the wave packets were propagating on geodesic trajectories, which are naturally the shortest paths, analogous to straight lines in flat geometry. On this background, we proposed shape-preserving accelerating beams in curved space [36], where the acceleration of beams is manifested in nongeodesic trajectories within the curved space. Thus far, however, no experiments have ever been carried out on shape-preserving accelerating beams in any curved-space system.

Here, we present the first experimental observation of accelerating beams in curved space. We demonstrate accelerating beams propagating within a thin spherical waveguiding layer, which forms a spherical-space setting for the EM wave. We find closed-form solutions for the wave equation in curved space that exhibit shape-preserving propagation along nongeodesic trajectories. Each of the lobes comprising these beams propagates along a different nongeodesic trajectory. These accelerating curved-space beams offer several features that are unique to curved-space settings: for example, they focus down in a scaled nongeodesic self-similar fashion, up to a plane where they break up, and subsequently undergo mirror reflection and start expanding again. This characteristic behavior occurs repeatedly as the accelerating (nongeodesic) beam travels around the surface of the sphere.

We begin by finding the accelerating shape-preserving beams propagating in a thin spherical dielectric shell. Since the problem is linear, we describe these solutions as a specific superposition of the eigenmodes of the EM field in this geometry. Consider the propagation of a monochromatic beam in a thin-film waveguide of curved spherical shape: a thin spherical dielectric shell. We begin with Maxwell's equations in spherical coordinates $(r, \theta, \phi)$, consider the wave polarized within the spherical shell (TM polarization), and derive the scalar wave equation for the electric field following Ref. [48]. When the radius of curvature is much larger than the wavelength, $2/R^2 \ll 1/\lambda^2$ (recalling that $2/R^2$ is a Gaussian curvature of a sphere), the longitudinal ($\phi$ component) of the TM field can be neglected [48]. The field is therefore scalar, $E = (E_r, E_\theta, E_\phi) = (0, \zeta, 0)$. We further assume that the spherical shell is thin and can support only one guided mode in the radial direction. This yields

$$-\nabla^2_\gamma \zeta - \partial_r^2 \zeta = k^2 \zeta, \quad (1)$$

where $\nabla^2_\gamma$ is the 2D Laplacian in the coordinates tangent to the surface, $\gamma$ is the metric of the surface, $\partial_r$ is the derivative with respect to the radial coordinate, and $k = k_0 n_0$, where $k_0$ is the free-space wave number and $n_0$ is the refraction index of the spherical shell. The metric of a sphere is

$$ds^2 = R^2 d\theta^2 + \cos^2(\theta) R^2 d\phi^2,$$
$$\gamma = \begin{pmatrix} R^2 & 0 \\ 0 & R^2 \cos^2(\theta) \end{pmatrix}, \quad (2)$$

where $R$ is the radius of the shell and $\theta = 0$ is on the equator. Substituting in Eq. (1) yields

$$-\frac{1}{R^2} \partial_\theta^2 \zeta + \frac{1}{R^2} \tan(\theta) \partial_\theta \zeta - \frac{1}{R^2 \cos^2(\theta)} \partial_\phi^2 \zeta - \partial_r^2 \zeta = k^2 \zeta. \quad (3)$$

Under the above assumptions, the electric field can be written as $\zeta = \psi(\theta, \phi) \xi(r)$, which decouples Eq. (3) into two equations,

$$\partial_r^2 \xi(r) + k_R^2 \xi(r) = 0, \quad (4)$$

$$-\frac{1}{R^2} \partial_\theta^2 \psi(\theta, \phi) + \frac{1}{R^2} \tan(\theta) \partial_\theta \psi(\theta, \phi)$$
$$- \frac{1}{R^2 \cos^2(\theta)} \partial_\phi^2 \psi(\theta, \phi) = k_T^2 \psi(\theta, \phi), \quad (5)$$

where $k_R$ and $k_T$ are the wave numbers in the radial and the transverse directions, respectively, such that $k^2 = k_R^2 + k_T^2$. Equation (4) describes the field dependence on the radial direction and Eq. (5) describes the tangential part. In the radial direction, the field is guided by the thin shell, whose guided modes are $\xi(r)$. For the tangential direction, we use separation of variables, defining $\psi(\theta, \phi) = X(\theta) Z(\phi)$, and derive two equations,

$$\partial_\phi^2 Z(\phi) + (k_\phi R)^2 Z(\phi) = 0, \quad (6)$$

$$-\frac{1}{R^2}\partial_\theta^2 X(\theta) + \frac{1}{R^2}\tan\theta \partial_\theta X(\theta) = \left(k_T^2 - k_\phi^2 \frac{1}{\cos^2\theta}\right) X(\theta), \quad (7)$$

where $k_\phi$ is the wave number in the $\phi$ direction ($|k_\phi| \leq |k_T|$). The solutions $X(\theta)$ of Eq. (7) are the associated Legendre polynomials $Y_l^m$, which serve as a complete orthogonal basis of eigenmodes for the EM field within the spherical shell. The solutions of Eq. (6) are forward and backward propagating waves. Henceforth, we consider only for the forward propagating solutions, as we are interested in a beam launched from $\phi = 0$. Getting back to the full solution, a propagating (nonevanescent) TM mode is given by $\zeta = X(\theta)Z(\phi)\xi(r) = Y_l^m(\theta, \phi)\xi(r)$, where the radial function $\xi(r)$ is the guided mode of the radial shell (acting as a dielectric waveguide), whose actual structure [51] bears no importance henceforth (except for setting the value of $k_T$) and $Y_l^m(\theta, \phi)$ is a spherical harmonic. The integer order and degree of $Y_l^m$, in Eqs. (6) and (7), are determined by the wave numbers $l = k_T R$ and $m = k_\phi R$, respectively. A general beam propagating in this setting can be described as a superposition of eignemodes,

$$\psi_l(\theta, \phi) = \sum_m A_m P_l^m(\cos\theta)e^{-im\phi}, \quad (8)$$

where $P_l^m(\cos\theta)$ is the associated Legendre polynomial and $A_m$ is the complex amplitude of the $m$th mode. These solutions are eigenmodes; that is, each one of them accumulates only phase as it evolves on the spherical shell. An example of the transverse structure of a mode is depicted in Fig. 1(a); this mode has two intense lobes and a weaker oscillating tail in between. Being an eigenmode, this mode is shape invariant; however, a superposition of modes does experience diffraction effects, as beams generally do. That is, an arbitrary initial beam $\psi_l(\theta, \phi = 0)$ experiences diffraction effects because its eigenmodes accumulate different phases as they evolve in $\phi$, according to Eq. (8). An example of such superposition of modes is the beam shown in Fig. 1(b), and its modal power spectrum (absolute values of the modal amplitudes) is shown in Fig. 1(e).

Next, we construct a localized beam that propagates along a nongeodesic trajectory in a self-similar fashion. A convenient way to construct such a beam is by truncating a single transverse eigenmode in the spirit of Ref. [52], which proposed a method of finding accelerating beams in flat space by truncating a high-order Hermite-Gauss beam. For the truncated beam to be approximately shape preserving, the truncation has to be done in such a way that the beam bears close resemblance to the truncated eigenmode. Under such conditions, the truncation of an eigenmode projects onto multiple eigenmodes, with the largest projection on the original eigenmode whose truncation forms the beam. Specifically in our problem, we find that truncating an eignemode in an exponential fashion produces a beam that is propagating self-similarly for a finite distance, whose extent is determined by the spatial truncation. Even more importantly, each of the individual lobes of this beam follows nongeodesic trajectories in a self-similar fashion, for a finite distance (with the following single exception: the eigenmode with $l = m$ has only a single lobe, and its trajectory follows a geodesic line).

An example of a beam formed by exponential truncation of the eigenmode $l_0 = 1000$, $m_0 = 950$ is shown in

$$\psi_{m_0,l_0}(\theta,\phi) = \sum_m e^{-(m-m_0)^2/N^2} P_{l_0}^m(\cos\theta) e^{-im\phi}. \quad (9)$$

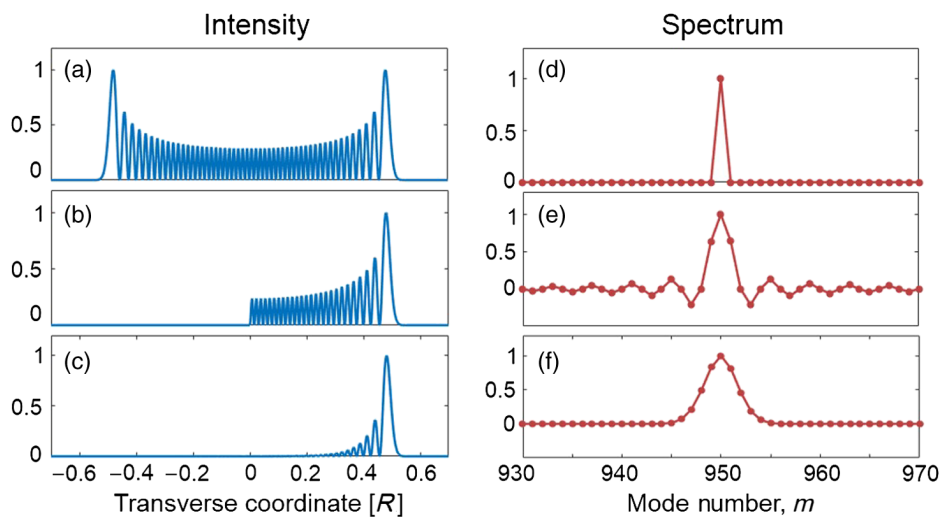

FIG. 1. (a)–(c) Calculated intensity cross sections of (a) a single transverse eigenmode (mode number $l = 1000$, $m = 950$) of the spherical shell, (b) the beam formed by abruptly cutting this mode, and (c) the shape-preserving accelerating beam constructed from a Gaussian superposition of modes of the same phase centered around $l_0 = 1000$, $m_0 = 950$; the lobes of this mode propagate in nongeodesic trajectories in a self-similar fashion. Experimentally, this beam can be generated by truncating the $l_0 = 1000$, $m_0 = 950$ eigenmode with an exponential density filter. The transverse coordinate is given in units of the radius $R$. (d)–(f) Corresponding spatial power spectra of the beams depicted in (a)–(c) as a function of mode number.

Fig. 1(c). This accelerating beam can be written as a Gaussian superposition of associated Legendre polynomials:

This construction yields accelerating beam with lobes that are evolving on different trajectories, whose parameters are set by $m_0$ and $l_0$. The integer $m_0$ is the degree of transverse mode that was truncated to form the accelerating beam. $m_0$ also defines the center of the modal spectrum of the beam, and can vary in the range $0 \leq m_0 \leq l_0$. The integer $l_0$ is a constant associated with the radial wave number ($k_R$) of the mode of the waveguide formed by the dielectric shell [defined by Eq. (4)], which yields $l_0 = R\sqrt{k^2 - k_R^2}$. For simplicity, we shall henceforth assume that the waveguiding shell is thin enough to support a single radial guided mode, with a propagation constant $k_R$, as schematically shown in Fig. 2(a). The summation over $m$ defines the modal spectrum of the accelerating beam. This construction yields a highly asymmetric beam: the field decays quickly to zero on one side of the main lobe, whereas the other side has multiple lobes propagating in different parallel trajectories (latitudes) on the surface of the sphere. This asymmetry in the beam structure is the curved-space manifestation of the intimate relation between Airy beams and caustics [1,52]. The trajectory of the accelerating beam (as a whole) changes according to the choice of $m_0$. Namely, each of the lobes of an accelerating beam is evolving on a trajectory of constant $\theta$; the higher the integer $(l_0 - m_0)$, the larger the curvature of this trajectory, and the closer it gets to the pole of the sphere. For a given radial mode $l_0$, the highest acceleration is obtained by truncating the transverse mode with the lowest $m_0$. An example of an accelerating beam constructed in this way—as a Gaussian superposition of modes—is shown in Fig. 1(c).

Next, we simulate the propagation of such accelerating beams, within the spherical shell forming the curved-space environment, Fig. 2(a). To do that, we develop a nonparaxial beam propagation method for EM waves propagating on the surface of the sphere. Figure 2(b) displays the propagation evolution of three different accelerating beams, highlighting that their trajectories deviate considerably from the geodesic line. The main lobe of each beam accelerates on a different nongeodesic trajectory marked by the red dashed lines. Each of these beams has additional lobes, located on a single side of their main lobe, while the field on the other side quickly decays to zero (forming a caustic). These sidelobes also evolve on nongeodesic lines, which are parallel to the trajectory of the main lobe of each of the beams, respectively. The accelerating trajectories with higher $l_0 - m_0$ have higher curvatures. Moreover, as $l_0 - m_0$ becomes larger, the transverse momentum carried by these beams increases, manifested by the increased density of the lobes for larger $l_0 - m_0$.

These beams have several interesting properties. Most notably, it is easy to see that the main lobe of these beams propagates on a nongeodesic trajectory, which is different from the trajectory of "center of mass" (see simulations in Figs. 2 and 3). To study this nongeodesic acceleration, we compare the trajectory of the accelerating main lobe of this beam with the trajectory of a Gaussian beam whose position and width is the same as of the main lobe of the accelerating beam, when both beams are launched from the same position (say, the $\phi = 0$ meridian). The trajectories are marked by the arrows in Fig. 3(a). The difference between the trajectories yields a cosine function, which, within the range drawn here, is well approximated by quadratic dependence on the propagation distance. Second, the transverse structure of the beams, $\psi_{l_0,m_0}(\theta, \phi = 0)$, flips every 180° of propagation, namely, $|\psi_{l_0,m_0}(\theta, \phi = 0)|^2 = |\psi_{l_0,m_0}(-\theta, \phi = \pi)|^2$, as shown in Fig. 3(b). This effect is due to the fact that the curvature of the spherical shells acts as a lens for the propagating beam: the propagation maps the beam to its modal spectrum after propagating a quarter of a sphere (at $\phi = \pi/2$), and then the beam is recovered mirror-reflected. Third, the direction of the main lobe relative to the geodesic direction (which is what the center of mass follows) can be varied by scaling the transverse shape of the beam. For a narrow launch beam, the direction of the main lobe is moving away from the geodesic line, similar to an Airy beam in flat space. But in addition, the curvature affects the propagation dynamics, acting like an effective potential. When the beam width is increased the trajectory of the main lobe is closer to the geodesic line [Fig. 3(c)], but never crosses over to the other side. Actually, in the limit of very broad lobes, the trajectories of the accelerating beams in curved space coincide with those in flat space, as one may expect. Altogether, in flat space, paraxial and nonparaxial, making the beam narrower always makes the trajectory steeper, but in curved space—as we notice here—what matters is the interplay between the width of the accelerating beam and the curvature of space which acts as an effective potential. The trajectory of the beam varies as a function of the width of its main lobe and as a function of the radius of the sphere. Acting in this fashion, the trajectory manifests the interplay between interference effects causing the acceleration and the curvature of the surface upon which the beam is propagating. Hence, unlike the Airy beam of flat space where the beam width uniquely defines the trajectory, here the acceleration trajectory is not unique: the same trajectory can be followed by beams propagating on spheres of different radii, by properly engineering the widths of its lobes. Thus, the trajectory of a shape-preserving accelerating beam on a sphere is affected not only by the width of its lobes (as all accelerating beams do) but also by the curvature of the surface upon which it is propagating.

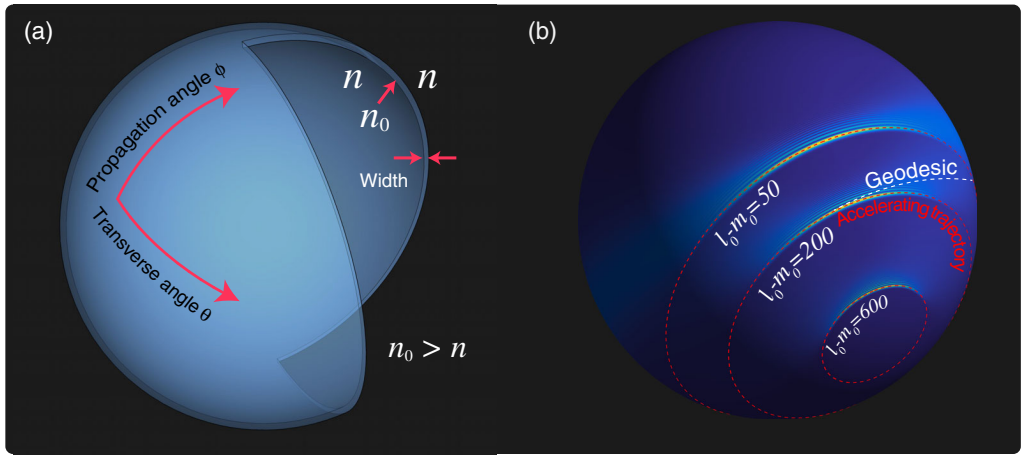

FIG. 2. (a) Sketch of a spherical shell with the coordinate system marked. The beams propagate in the $\phi$ direction and their transverse structure, shown in Fig. 1(c), extends in the $\theta$ direction. The thickness of the spherical shell is marked by the red arrows. In this region the refractive index $n_0$ is higher than in the surrounding medium $n$, and therefore the mode is confined within the shell. The structure of the beams in the radial direction is dictated by Eq. (4). (b) Three accelerating beams that follow three different nongeodesic trajectories. A beam with a larger $l_0 - m_0$ index follows an acceleration trajectory with a higher curvature. The larger $l_0 - m_0$, the higher the density of the lobes, which corresponds to higher transverse momentum. The red dashed lines are the acceleration trajectories of the beams, while the white dashed line shows the geodesic trajectory, which is the trajectory that a Gaussian beam propagating on a sphere would follow. Notice that each of these beams forms a caustic: the field on one side of the main lobe quickly decays to zero, while the other side of the beam consists of multiple lobes propagating in parallel on nongeodesic trajectories.

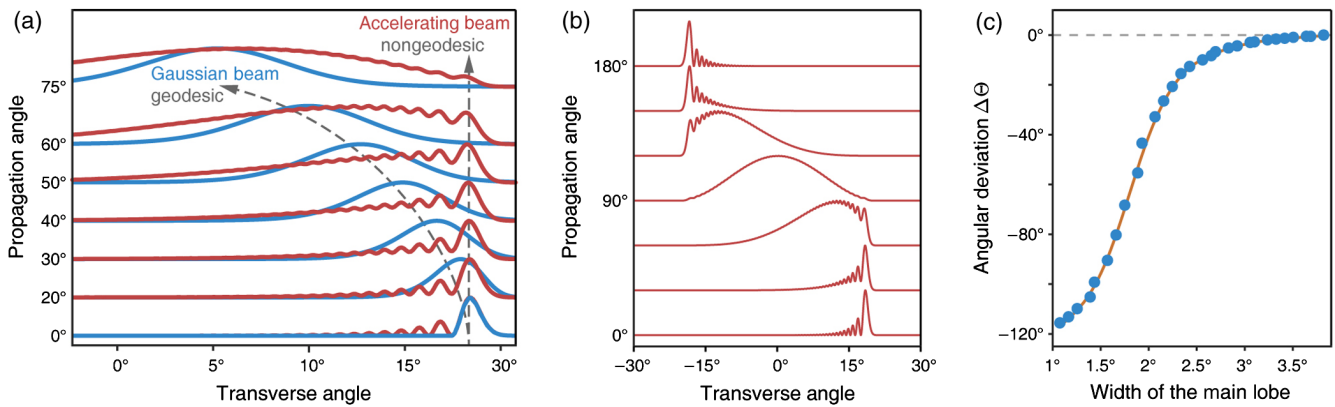

FIG. 3. (a) Simulated propagation on the surface of a sphere of radius $R = 1$ cm, for the accelerating beam of $l_0 = 1000$, $m_0 = 950$ (red lines) whose structure is shown in Fig. 2(b), and for a beam composed of the main lobe only (blue lines). The accelerating beam follows a nongeodesic path, whereas the isolated main lobe follows the expected geodesic line. The trajectories are marked by gray arrows. (b) Propagation evolution of the accelerating beam over a hemisphere (180° on the sphere). The spatial profile of the beam flips after propagating over half the sphere. (c) Angle difference $\Delta\Theta$ between the geodesic line and the trajectory of the main lobe of the accelerating beam of (a), as function of the main lobe width, in angular units (deg). When the lobes are very narrow, $\Delta\Theta$ has a large absolute value and therefore the acceleration is highly visible, whereas when the width of the lobes is increased, the lobes tend to approach geodesic trajectories. The same principle holds if the size of the input beam is kept fixed but the radius of the sphere is varied. Namely, increasing the radius of the shell has the same effect on the accelerating beam as decreasing the beam's transverse size.

Finally, we demonstrate these shape-preserving accelerating (nongeodesic) beams in experiments, using the setup sketched in Fig. 4(a). Our thin spherical shell is a crown-glass hemisphere of 3 cm radius and 550 $\mu$m thickness [Fig. 4(b)]. It is the shell of an incandescent light bulb cut in half. We form the structured beam by reflecting a 532 nm laser beam off a spatial light modulator that generates the desired Gaussian modal superposition forming the accelerating beam. The structure of the accelerating beam in real space is formed at the focal plane of the cylindrical lens. This structured beam is coupled into the waveguiding layer of the spherical shell. The surface of the shell is painted so as to produce scattering, to make the propagation of the beam visible. The scattered light is image onto a CCD camera through a microscope objective [Fig. 4(c)].

The experimental results are shown in Fig. 5(a). First, we launch the accelerating beam, which is a Gaussian superposition of eigenmodes centered around the mode with

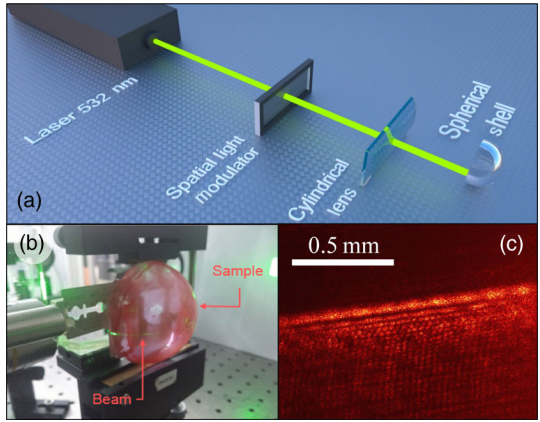

FIG. 4. (a) Experimental setup. A collimated laser beam ($\lambda = 532$ nm) is reflected off a spatial light modulator to generate the desired 1D spectrum (Gaussian modal superposition) of the beam, which is then Fourier transformed by a cylindrical lens to form the predesigned shape-preserving accelerating beam. This beam is then coupled to the hemispherical thin glass shell (3 cm radius, 550 $\mu$m width). (b) Experimentally observed propagation trajectory of the beam accelerating on the surface of a spherical shell. The propagation inside the shell is visible as a green beam thanks to the scattering from the red paint of the shell. (c) Zoom-in photograph of the lobes of the accelerating beam, as it propagates within the spherical shell.

$l_0 = 567\,000$ and $l_0 - m_0 = 50$; the value of $l_0 \approx 2\pi R n_0/\lambda$ is defined by the radius of the sphere and the wavelength. In the radial direction, this mode is approximately solely the lowest radial mode, the solution of Eq. (4). We observe the expected behavior during propagation—the beam propagates in a nongeodesic trajectory in a self-similar fashion as presented in Fig. 5(a). Second, for comparison, we also launch a Gaussian beam, which is basically the isolated main lobe of the accelerating beam (as the rest of the beam is truncated). As shown in Fig. 5(a), the Gaussian beam experiences diffraction broadening and follows a geodesic trajectory, whereas the accelerating beam maintains its shape for a considerable distance (at least up to 50° of propagation), and its main lobe clearly follows a nongeodesic trajectory up to 75° of propagation. The other lobes of the accelerating beam also evolve on nongeodesic lines, although they seem to maintain their shape for distances shorter than that of the main lobe [Fig. 5(a)]. With the curved-space setting used in this experiment (the light bulb), it is not possible to observe the beam profile for longer propagation distances since the amplitude of the beam decreases due to propagation losses (mostly scattering from surface roughness). To summarize the experimental results, when we launch only the main lobe, the beam evolves on a geodesic trajectory [blue profiles in Fig. 5(a)], essentially conforming to the propagation of a Gaussian beam on a sphere. In order to observe acceleration, we add the rest of the lobes, and observe that the main lobe is now following a trajectory that is very different than the geodesic line [red profile in Fig. 5(a)]. By measuring the difference between these trajectories, we experimentally evaluate the deviation from the geodesic line, which gives the spatial acceleration of the accelerating beam. As shown is Fig. 5(b), the observed experimental deviation from the geodesic line is quadratic with propagation distance, as predicted by the simulation.

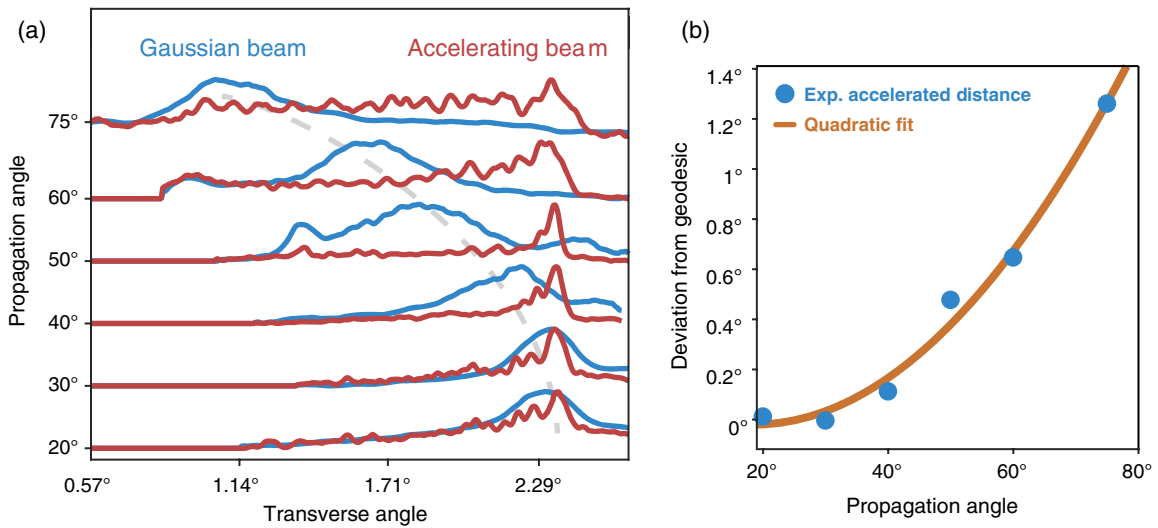

FIG. 5. (a) Experimentally measured intensity profiles of the self-similar accelerating beam (red lines) and of the single-lobe beam (blue lines), at various propagation stages on the spherical surface. The single-lobe beam is generated by taking just the main lobe of the accelerating beam at the input plane. This beam experiences diffraction broadening while propagating on a geodesic line. On the other hand, the accelerating beam is shape preserving up to 50°, while it propagates on a nongeodesic trajectory. This is clearly observed by the increasing shift between the main lobe of the accelerating beam and the single-lobed beam. (b) The experimentally measured angular difference between the single-lobed beam and the main lobe of the accelerating beam, as a function of angular propagation distance on the spherical surface. The parabolic relation of this distance, and the fact that our structured beam maintains its shape for a large distance, experimentally confirm the accelerating shape-preserving nature of our beam.

In conclusion, we find shape-preserving accelerating beams on spherical surfaces. The lobes of these beams propagate along nongeodesic trajectories in a self-similar fashion. We simulate their propagation and compare it to the propagation of beams that follow geodesic trajectories. Finally, we observe these accelerating beams in experiments. To our knowledge, this is the first observation of shape-preserving accelerating beams in curved space. In this vein, we envision shaping and manipulating the propagation of beams in curved-space environments, and use such beams to manipulate nanoparticles in nonplanar settings such as microchannels, blood vessels, etc. Likewise, accelerating plasmonic beams in curved-space settings can be designed for metallic bodies such as gold spheres and ellipsoids, and used for various applications that require transfer of power from a broad area to a small region upon a nonplanar surface. Finally, designing shape-preserving beams in curved space will be very helpful in emulating 3D general relativity phenomena in optical settings, expanding the vision of Ref. [46] into 3D.

## ACKNOWLEDGMENTS

This work was funded by the Israel Science Foundation, by the Israel Science Ministry, and by the U.S. Air Force Office of Scientific Research. The research leading to these results has received funding from the European Union's–Seventh Framework Programme (FP7/2007-2013) under Grant Agreement No. 629114 MC–Structured Light.

A. P. and M. A. B. contributed equally to this work.

PATSYK, BANDRES, BEKENSTEIN, and SEGEV        PHYS. REV. X 8, 011001 (2018)